\documentclass[letterpaper, 10 pt, conference]{ieeeconf}  

\IEEEoverridecommandlockouts                              

\usepackage{graphics} 
\usepackage{epsfig} 
\usepackage{mathptmx} 
\usepackage{times} 
\usepackage{cite}
\usepackage{amsmath,amssymb,amsfonts}
\usepackage{algorithmic}
\usepackage{graphicx}
\usepackage{textcomp}
\usepackage{multirow}
\usepackage{array}
\usepackage{tikz}

\title{\LARGE \bf
 Decoding of Grasp Motions from EEG Signals \\Based on a Novel Data Augmentation Strategy  
}

\author{Jeong-Hyun Cho, Ji-Hoon Jeong, and Seong-Whan Lee, \textit{Fellow}, \textit{IEEE}
\thanks{*This research was partly supported by an Institute of Information \& Communications Technology Planning \& Evaluation (IITP) grant, funded by the Korean government (No. 2017-0-00432, Development of Non-Invasive Integrated BCI SW Platform to Control Home Appliances and External Devices by User’s Thought via AR/VR Interface) and partly funded by an Institute of Information \& Communications Technology Planning \& Evaluation (IITP) grant funded by the Korean government (No. 2017-0-00451, Development of BCI based Brain and Cognitive Computing Technology for Recognizing User’s Intentions using Deep Learning).}
\thanks{J.-H. Cho and J.-H Jeong are with the Department of Brain and Cognitive Engineering, Korea University.
        }%
\thanks{S.-W. Lee is with the Department of Artificial Intelligence, Korea University, 145, Anam-ro, Seongbuk-gu, Seoul 02841, Republic of Korea (corresponding author: sw.lee@korea.ac.kr). }
\thanks{Accepted for publication at EMBC 2020. {\textcopyright}2020 IEEE}
}

\begin{document}

\maketitle
\thispagestyle{empty}
\pagestyle{empty}

\begin{abstract}

Electroencephalogram (EEG) based brain-computer interface (BCI) systems are useful tools for clinical purposes like neural prostheses. In this study, we collected EEG signals related to grasp motions. Five healthy subjects participated in this experiment. They executed and imagined five sustained-grasp actions. We proposed a novel data augmentation method that increases the amount of training data using labels obtained from electromyogram (EMG) signals analysis. For implementation, we recorded EEG and EMG simultaneously. The data augmentation over the original EEG data concluded higher classification accuracy than other competitors. As a result, we obtained the average classification accuracy of 52.49($\pm$8.74)\% for motor execution (ME) and 40.36($\pm$3.39)\% for motor imagery (MI). These are 9.30\% and 6.19\% higher, respectively than the result of the comparable methods. Moreover, the proposed method could minimize the need for the calibration session, which reduces the practicality of most BCIs. This result is encouraging, and the proposed method could potentially be used in future applications such as a BCI-driven robot control for handling various daily use objects.
\newline

\indent \textit{Clinical relevance}— This study suggests a method to improve the decoding performance and practicality of motor intention-based BCIs so the patient could control a neuroprosthesis with sufficient accuracy. 
\end{abstract}

\section{INTRODUCTION}

Electroencephalogram (EEG)-based brain-computer interfaces (BCIs) represent a new way to translate human intentions into external device commands via brain activity, and many BCI systems have been developed to combine with many different types of applications such as upper limb prosthesis \cite{jeong2020decoding, agashe2015global, jeong2020brain} and have many other possibilities \cite{lafleur2013quadcopter, chen2016high, ding2013changes, zhu2016canonical}. Motor imagery (MI) is one of the most common EEG modalities, and recent studies on an MI-based BCI system have proved their potential for facilitating the reliable control of external devices such as robotic arms and gloves \cite{schwarz2017decoding, ofner2017upper, ang2016eeg, agashe2015global}. Meanwhile, motor execution (ME), which is induced by actual muscle activity, is also widely used in the study of motor-related BCIs because more distinct EEG signals can be obtained in many scenarios \cite{schwarz2017decoding, ofner2017upper, jeong2019trajectory}. These advances in ME- and MI-based BCIs, which have allowed people to communicate with external devices through thoughts rather than the peripheral nervous system, have demonstrated an impressive capacity for enhancing human performance.

Grasping is associated with more dynamic brain activity than movements of other extremities because a large area of the human brain's motor cortex is allocated to controlling hands \cite{veres2017modeling, yun2012hand, robinson2013eeg, jeong2020brain, jeong2018decoding}. From this dynamic background activity, we can identify EEG signals of various strengths and sources. Many recent works have managed to successfully decode EEG motor intentions related to movements of large body parts such as arms and feet with a diverse range of methods and approaches, and some of them have inspired our present research. We tried to extend this inspiration toward decoding complex grasp motions rather than toward other extremities.

The recent BCI studies were carefully executed regarding the user's motor intention, but a significant problem they encountered was difficulty training classifiers due to small dataset \cite{zhang2019novel, kwak2017convolutional, park2017filter, zhang2018temporally}. To overcome this limitation, some researchers proposed the idea of extracting various features from the limited amount of data. They tried to obtain distinct features in the spatial, temporal, and spectral domains. Also, using regularization to spatial filters was an option to maximize the decoding performance under the limitation of the small dataset. Of course, increasing the amount of raw data using data augmentation methods is a valid option \cite{zhang2019novel}, but applying the augmentation method to EEG data was implemented on a limited basis until recently because it was challenging to find proper augmentation methods for EEG signals.

Therefore, the objective of this study is to suggest a novel data augmentation strategy on the original EEG data by mixing segmented EEG data, which are from different trials and classes. Using the labels obtained by decoding electromyogram (EMG) signals allows this approach. The proposed method, named label-based data augmentation, successfully creates a large amount of augmented EEG data to extract features and train classifiers as the original data could do. As a result, we proved the feasibility of classifying five complex grasp motions in the right hand from EEG signals with the proposed method in both ME and MI paradigms. Using this method, we improved classification accuracy, so it will be used for further BCI applications, such as controlling a robotic hand. By running sufficient experimental trials and data analysis, we could construct a robust decoding model based on our proposed method.

\begin{figure}[t!]
\begin{center}
\includegraphics[width=0.80\columnwidth]{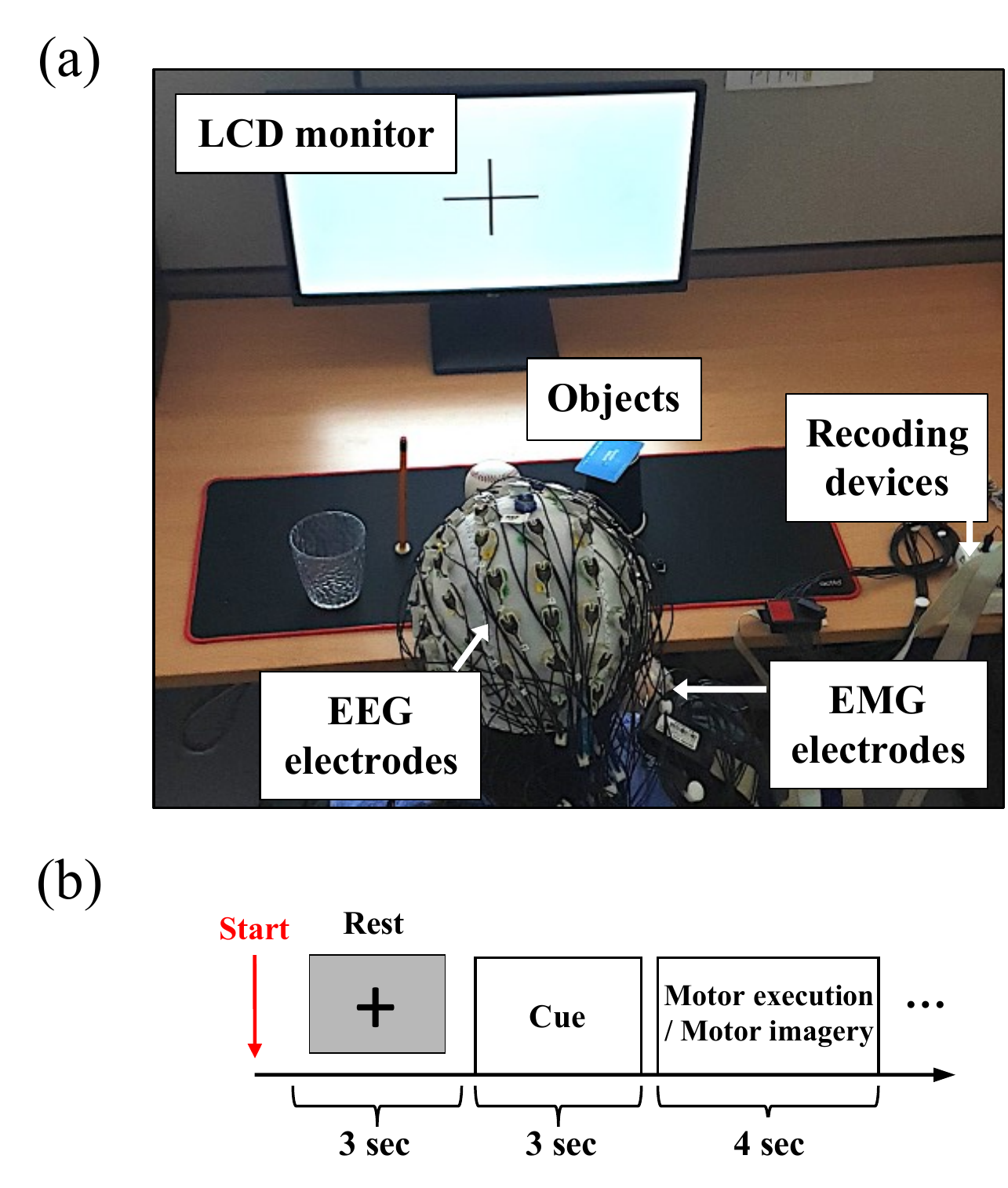}
\end{center}
\caption{Description of experimental setup and protocol for EEG and EMG signals acquisition induced by five grasp motions. a) Experimental environment and setup. b) Experimental protocol. The subjects perform the motor execution (ME) and motor imagery (MI) after the visual cue.}
\label{fig1}
\end{figure}

\section{Materials and Methods}

\subsection{Participants}
Five healthy subjects with no history of neurological disease were recruited for the experiment (S1-–S5; ages 25-–34; five men; all right-handed). This study was reviewed and approved by the International Review Board, at Korea University [1040548-KU-IRB-17-172-A-2], and written informed consent was obtained from all participants before the experiments. 

\subsection{Experimental Setup}
During a session of the experimental protocol, the subjects sat in front of a 24-inch LCD monitor screen, in a comfortable chair. Fig. 1 (a) indicates the experimental setup and the environment during the entire session. In the experiment, the subjects were asked to perform five different grasp actions or imagery, as shown in Fig. 1 (b). The location of the object setup was randomly changed to reduce the effect of artifacts. Each subject performs 250 trials (50 trials $\times$ 5 classes) per session.   

\subsection{Data Acquisition}
EEG data were collected at 1,000 Hz using 20 Ag/AgCl electrodes (FC1--6, C1--6, Cz, CP1--6, and CPz) in the 10/20 international system via BrainAmp (BrainProduct GmbH, Germany) \cite{jeong2020brain}. The 20 channels were located only on the motor cortex to ensure that the recorded EEG signals corresponded to the motor-related potentials under ME and MI paradigms.

EMG signals were recorded using 5 Ag/AgCl electrodes and a digital amplifier, which is the same equipment used to record EEG signals. The details of the related muscles are as follows: extensor carpi ulnaris (CH1), extensor digitorum (CH2), flexor carpi radialis (CH3), and flexor carpi ulnaris (CH4). The last electrode was placed near to the elbow, which is a non-muscle movement area, for a reference signal. The signals were first processed by a 60 Hz notch and filtered by a [10--500] Hz band-pass filter \cite{trigili2019detection, li2017motion, yao2013combining}.

\subsection{Data Analysis}
The entire process of decoding EEG signals induced by the subject performing grasp actions and imagery was described in Fig. 2. We followed a conventional processing flow for EEG and EMG signals bandpass filtering, channel selection, feature extraction, and feature selection \cite{jeong2018decoding, suk2014predicting, kwak2015lower, won2015effect, ang2012filter}. Also, median filter was applied to the EEG data in order to concatenate segmented data from different sources smoothly. In the case of EEG signals decoding, we extracted spatial features based on common spatial patterns (CSP) from the preprocessed signals using the first and the last three CSP filters \cite{ramoser2000optimal, kim2016commanding, suk2011subject, park2017filter, zhang2018temporally}. After the feature extraction, we trained classifiers with the one-versus-all method for multiclass classification. For classification, we adopted the support vector machine (SVM). Every classifier was trained with the augmented data and tested with pre-selected trials of the original data.

\begin{figure}[t!]
\centering
\includegraphics[width=0.85\columnwidth]{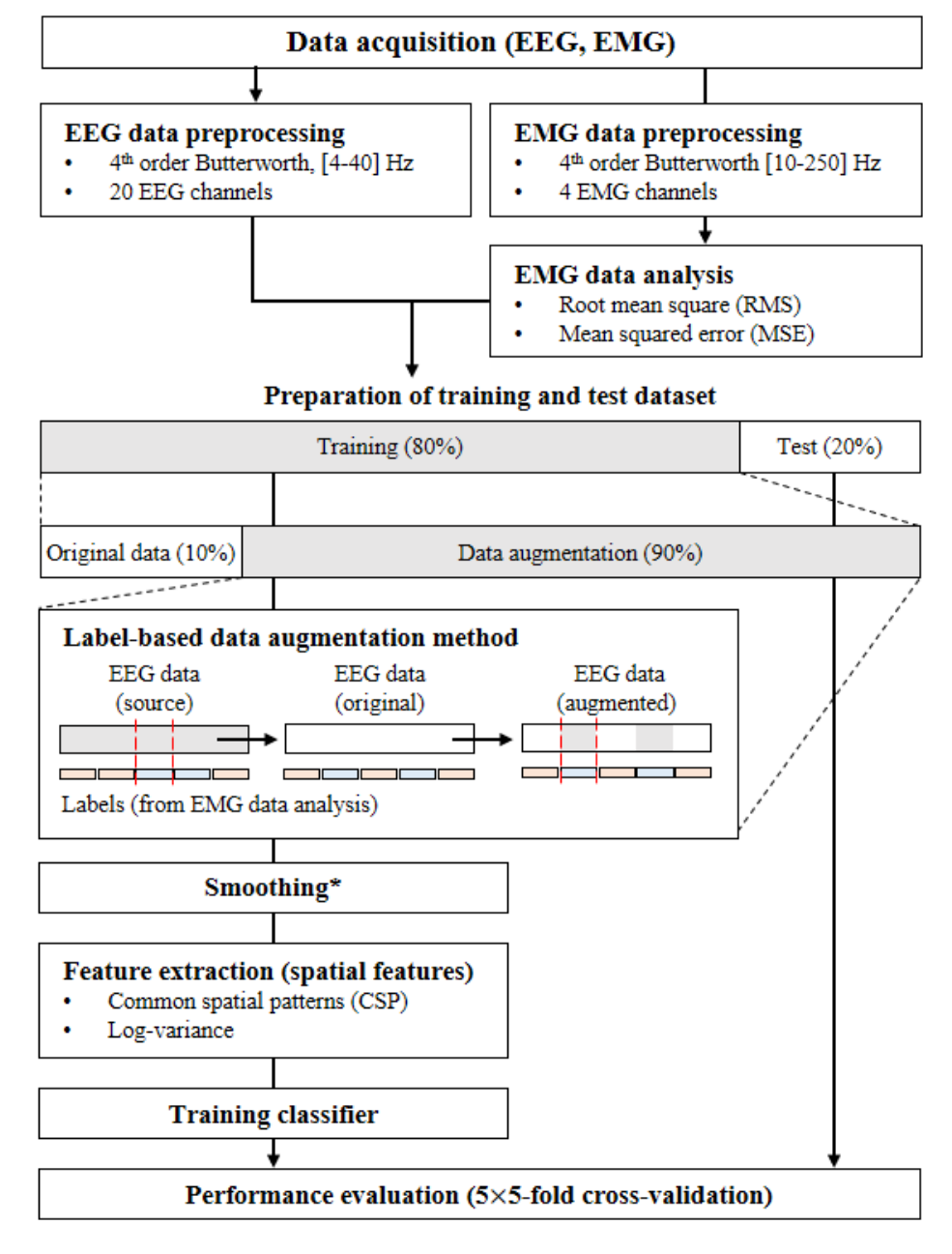}
\caption{Overall flowchart of the proposed data augmentation method using labels based on EMG analysis. $^*$Smoothing returns the filtered signal in which discontinuous sections have been removed.}
\label{fig2}
\end{figure}

The proposed method divides data (4 sec) of a single trial into fifteen segments and adds labels obtained by decoding EMG signals. The 500 ms sliding window with a 250 ms step size was applied to make the segments. The same sliding window was applied to the EMG signals as well to match the time stamp to the segmented EEG data. After creating the segments, we calculate the root mean square (RMS) value on the segmented EMG data \cite{li2017motion, trigili2019detection}. The labels denote which muscles were activated in a specific time interval based on calculating mean squared error (MSE). After this process, we can prepare EEG data, including labels. The labeled EEG segments from all classes are stored in advance, and we call this a segmented data bank. In the last step, we randomly switched the original segments of EEG data into the segments, which are from the data bank with the same label (the smallest MSE). As a result, the original EEG data that was mixed with various segments extracted randomly in different trials and classes becomes augmented EEG data. According to our test, switching 60\% of the data segments showed the highest performance for classification accuracy improvement, so we used the ratio in this study.

\begin{table}[t!]
\centering
\caption{Classification results of proposed and comparable methods in ME and MI paradigms}
\normalsize
\label{table1}
\renewcommand{\arraystretch}{1.1}
\resizebox{\columnwidth}{!}{%
\begin{tabular}{cccccc}
\hline
\multirow{2}{*}{\textbf{Subject}} & \multirow{2}{*}{\textbf{Paradigm}} & \multicolumn{4}{c}{\textbf{Accuracy (\%)}} \\ \cline{3-6} 
 &  & \textbf{CSP} & \textbf{CSP$_{DA}$} & \textbf{FBCSP} & \textbf{FBCSP$_{DA}$} \\ \hline
\multirow{2}{*}{\textbf{S1}} & \textbf{ME} & 37.43 & 44.73 & 41.27 & 48.03 \\
 & \textbf{MI} & 24.83 & 24.43 & 26.03 & 40.75 \\
\multirow{2}{*}{\textbf{S2}} & \textbf{ME} & 40.47 & 52.81 & 48.27 & 61.67 \\
 & \textbf{MI} & 28.23 & 35.67 & 41.51 & 41.73 \\
\multirow{2}{*}{\textbf{S3}} & \textbf{ME} & 35.87 & 46.32 & 39.67 & 47.53 \\
 & \textbf{MI} & 30.11 & 38.05 & 36.93 & 40.37 \\
\multirow{2}{*}{\textbf{S4}} & \textbf{ME} & 39.17 & 44.74 & 41.53 & 43.21 \\
 & \textbf{MI} & 29.50 & 32.37 & 29.73 & 34.87 \\
\multirow{2}{*}{\textbf{S5}} & \textbf{ME} & 43.07 & 47.63 & 45.23 & 62.03 \\
 & \textbf{MI} & 27.77 & 38.13 & 36.61 & 44.07 \\ \hline
\multicolumn{1}{l}{\multirow{2}{*}{\textbf{Mean($\pm$Std)}}} & \textbf{ME} & 39.20($\pm$2.78) & 47.25($\pm$3.34) & 43.19($\pm$3.49) & \textbf{52.49($\pm$8.74)} \\
\multicolumn{1}{l}{} & \textbf{MI} & 28.09($\pm$2.05) & 33.73($\pm$5.70) & 34.16($\pm$6.19) & \textbf{40.36($\pm$3.39)} \\ \hline
\multirow{2}{*}{\textbf{p-value}} & \textbf{ME} & \textless 0.01 & -- & 0.024 & -- \\
 & \textbf{MI} & 0.043 & -- & 0.064 & -- \\ \hline
\end{tabular}%
}
\end{table}

\begin{figure}[t!]
\centering
\includegraphics[width=\columnwidth]{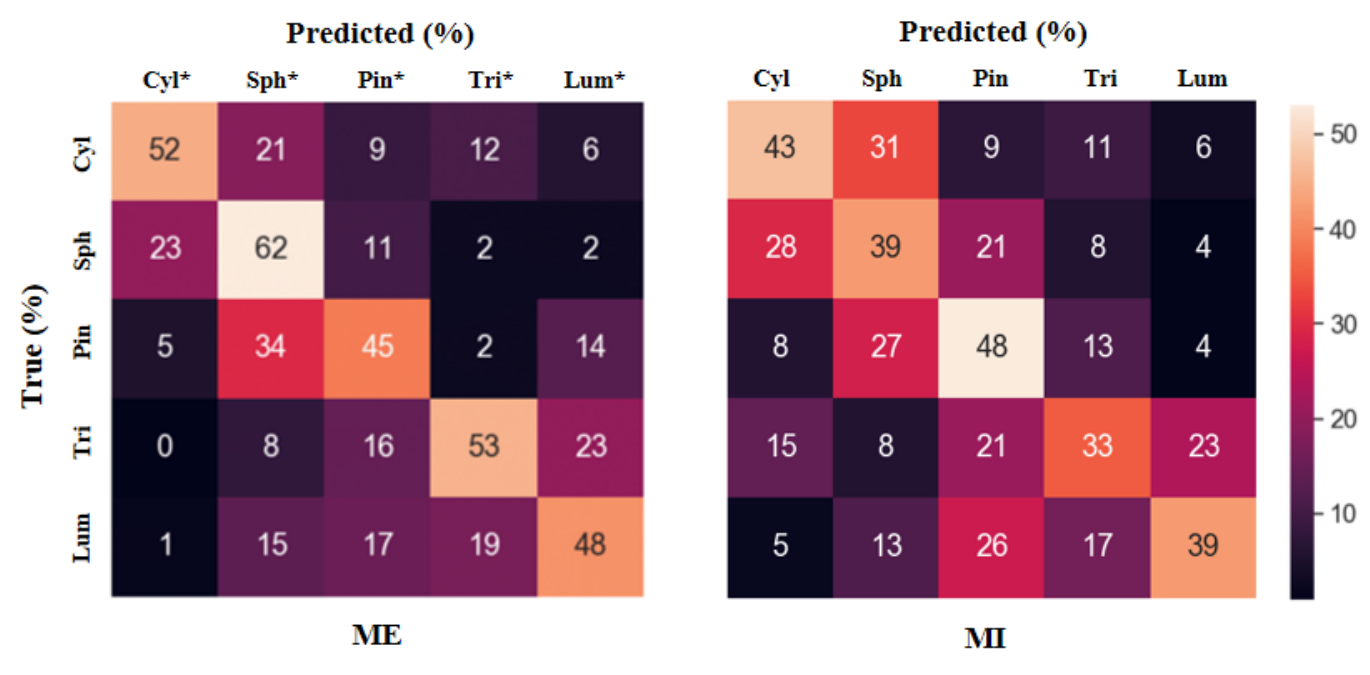}
\caption{Confusion matrices of average accuracy (FBCSP$_{DA}$) in ME and MI paradigms, respectively. $^*$Cyl: cylindrical grasp, $^*$Sph: spherical grasp, $^*$Pin: pinch grasp, $^*$Tri: Tripod grasp, $^*$Lum: lumbrical grasp.}
\label{fig3}
\end{figure}

\section{Results}
The proposed method improves the overall classification performance of the BCI system. We compared four methods named CSP, CSP$_{DA}$, FBCSP, and FBCSP$_{DA}$ to confirm the effectiveness of the label-based data augmentation, as shown in Table I. The filter bank CSP (FBCSP) usually shows higher performance than the standard CSP approach in other BCI studies \cite{jeong2018decoding}. The CSP and FBCSP denote that the classification was performed based on the trained classifiers using spatial features extracted by CSP and FBCSP without data augmentation. The CSP$_{DA}$ and FBCSP$_{DA}$ represent the classification using the augmented data. We used the same test data to compare the classification performance of the four methods. Our methods using the label-based augmentation showed a 8.05\% increase in classification accuracy compared to the CSP and a 9.30\% improvement compared to the FBCSP in ME. At the same time, the proposed method improves the decoding performance in the MI paradigm. It showed an 5.64\% and a 6.19\% improvement compared to CSP and the FBCSP, respectively. 5$\times$5-fold cross-validation was used, and also the test dataset was separated before the data augmentation process was executed. 

According to the results of this comparison, we can conclude that the proposed method improves the classification performance for the ME-BCIs and MI-BCIs. Of course, we also confirmed a reduction of classification accuracy in certain subjects, but the average classification accuracy was improved for the overall group. Not only does the proposed method lead to an improvement in classification accuracy, but it also brought an additional advantage to developing practical BCI systems: the proposed data augmentation showed stable classification performance, even with a minimized calibration session. Besides, the classification accuracy presented in confusion matrices shows even classification results in each class (Fig. 3).

\begin{figure}[t!]
\centering
\includegraphics[width=0.90\columnwidth]{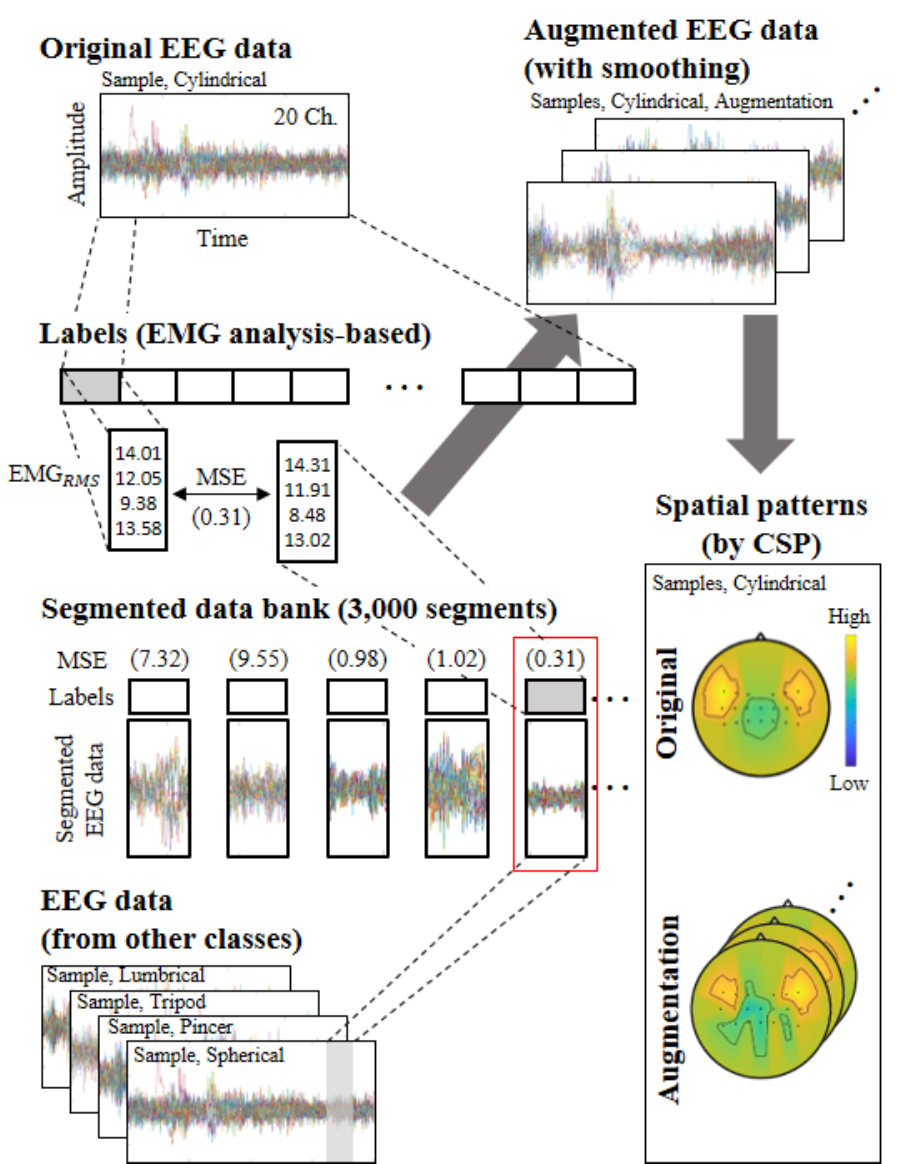}
\caption{Illustration describing data augmentation method using segmented EEG data. Each plot shows the normalized signals on 20 EEG channels (samples of 'cylindrical' class, subject S5). One trial is divided into 15 labeled segments, and 80\% of all trials (200 trials per subject) were used to create 3,000 segments. CSPs also presented for comparison.}
\label{fig4}
\end{figure}

\section{Discussion and Conclusion} 
Fig. 4 shows the details of the label-based data augmentation process. We plotted the normalized EEG signals to see the similarity between the original and augmented data. We think that the similarity of the augmented data is sufficient compared with the original data because we obtained realistic CSP pattern from the augmented data and also the classifiers trained with augmented data shows improved classification performance to the real EEG data which was used as the test dataset. In the MI paradigm, the proposed method showed less improvement in classification accuracy than in the ME because we could not obtain the corresponding EMG signals while the subjects performed MI. Therefore, we recalled the existing labels from the ME decoding process and then applied it based on an assumption that subjects imagined activities in a similar order to actual movements.

In conclusion, the proposed method suggested a novel approach to decode EEG signals to classify complex grasp motions. This method has the potential to achieve additional improvements in classification accuracy and practicality of use by preparing a complemented data acquisition environment and using advanced learning methods, such as deep learning, that were not considered confidently due to the needs of a large amount of training data \cite{tabar2016novel, kwak2017convolutional, uktveris2017application, tayeb2019validating}.
   
\section*{Acknowledgment}
The authors thank to K.-H. Shim, B.-H. Kwon, B.-H. Lee, D.-Y. Lee and D.-H. Lee for help with the database construction and useful discussions of the experiment. 

 
\bibliographystyle{IEEEtran}
\bibliography{ref.bib}

\end{document}